%% file: rev5.tex
\documentclass[fleqn,usenatbib]{mnras}

\input{package}     
\input{definitions} 
\input{abbreviation}
\usepackage{color}

\definecolor{awf}{RGB}{125,25,100}

\title[Ly$\alpha$ emission from galaxies in the EoR]{Ly$\alpha$ emission from galaxies in the Epoch of Reionization}

\author[C. Behrens et al.]{
C. Behrens,$^{1}$\thanks{\href{mailto:christoph.behrens@sns.it}{christoph.behrens@sns.it}},
A. Pallottini,$^{1,2,3,4}$,
A. Ferrara$^{1,5}$,
S. Gallerani$^{1}$, 
L. Vallini$^{6,7}$\\
$^{1}$Scuola Normale Superiore, Piazza dei Cavalieri 7, I-56126 Pisa, Italy\\
$^{2}$Centro Fermi, Museo Storico della Fisica e Centro Studi e Ricerche ``Enrico Fermi'', Piazza del Viminale 1, Roma, 00184, Italy\\
$^{3}$Cavendish Laboratory, University of Cambridge, 19 J. J. Thomson Ave., Cambridge CB3 0HE, UK\\
$^{4}$Kavli Institute for Cosmology, University of Cambridge, Madingley Road, Cambridge CB3 0HA, UK\\
$^{5}$Kavli IPMU, The University of Tokyo, 5-1-5 Kashiwanoha, Kashiwa 277-8583, Japan\\
$^{6}$Leiden Observatory, Leiden University, P.O. Box 9513, NL-2300 RA Leiden, The Netherlands\\
$^{7}$Nordita, KTH Royal Institute of Technology and Stockholm University, Roslagstullsbacken 23, SE-10691 Stockholm, Sweden
}

\date{Accepted 2019 April 04. Received 2019 April 02; in original form 2018 August 30}

\pubyear{2019}

\begin{document}
\label{firstpage}
\pagerange{\pageref{firstpage}--\pageref{lastpage}}
\maketitle

\begin{abstract}
The intrinsic strength of the \lyman\ line in young, star-forming systems makes it a special tool for studying high-redshift galaxies. However,  interpreting observations remains challenging due to the complex radiative transfer involved. Here, we combine state-of-the-art hydrodynamical simulations of \quotes{\althaea}, a prototypical Lyman Break Galaxy (LBG, stellar mass $M_{\star}\simeq 10^{10}\msun$) at $z=7.2$, with detailed radiative transfer computations of dust/continuum, \CII 158 $\mu$m, and \lyman\ to clarify the relation between the galaxy properties and its \lyman\ emission. 
\althaea\ exhibits low ($f_\alpha< 1$\%) \lyman\ escape fractions and Equivalent Widths, EW $\lesssim 6$ \AA{} for the simulated lines of sight, with a large scatter. The correlation between escape fraction and inclination is weak, as a result of the rather chaotic structure of high-redshift galaxies. Low $f_\alpha$ values persist even if we artificially remove neutral gas around star forming regions to mimick the presence of \HII\ regions.  The high attenuation is primarily caused by dust clumps co-located with young stellar clusters. We can turn \althaea\ into a Lyman Alpha Emitter (LAE) only if we artificially remove dust from the clumps, yielding EWs up to $22$ \AA{}. Our study suggests that the LBG-LAE duty-cycle required by recent clustering measurements poses the challenging problem of a dynamically changing dust attenuation.
Finally, we find an anti-correlation between the magnitude of \lyman --\CII\ line velocity shift and \lyman\ luminosity.
\end{abstract}




\section{Introduction}
Understanding the formation and evolution of young galaxies via the \lyman\ line produced by massive stars has become a standard tool of extragalactic observations. 
However, such a strategy is still affected by a number of uncertainties that can be ultimately associated with the resonant nature of this line, and that result in a complex relation between gas distribution, kinematics, and the observables.

This complexity fostered a large number of attempts to simulate \RT\ in galaxies, using both simplified - but yet successful - models \citep{Verhamme2006,Dijkstra2006,Behrens2014b} and more detailed models that follow the \RT\ in an isolated disk galaxy \citep{Verhamme2012,Behrens2014a}, or even galaxies in their cosmological environments \citep{Laursen2010,Gronke2017}.

While considerable progress has been made, many puzzles remain. An outstanding one concerns the decreasing abundance of \lyman\ emitting galaxies towards high-redshift and in the Epoch of Reionization (EoR). The luminosity function of \LAEs\ shows only mild evolution between redshift 2 and 6, but rapidly declines above $z=6$ \citep{Pentericci2011,Stark2010,Hayes2011,Schenker2012,Ono2012}. In particular, the \LAEs\ abundance drops more strongly than the UV luminosity function. 

Several proposals have been made to explain this discrepancy. One possibility is related to the evolution of the neutral hydrogen fraction in the intergalactic medium (IGM), which might rapidly increase at $z>6$ as a result of the cosmic reionization process. The problem with this explanation is that it requires a dramatic change of the IGM neutral fraction $x_{HI}$ of $\Delta x_{HI} \sim 0.5$ in a short redshift range, $\Delta z = 1$ \citep{Pentericci2011,Ono2012,Schenker2012,Dijkstra2011}, requiring both a late and very rapid reionization. While the latest Planck results \citep{Planck2018} do allow for a late reionization scenario, the rapidness of the transition is hard to model. However, e.g. \cite{Mesinger2015} find that a joint evolution of the ionization field and the neutral fraction is moderately consistent with the data. Other possible explanations of the vanishing of \LAEs\ focus on changes in the circumgalactic medium (CGM). \citet{Bolton2013,Weinberger2018}  argue that in the final stages of reionization, a decrease in the mean free path length of ionizing photons lowers the transmissivity of the CGM to \lyman\ photons; \citet{Sadoun2017} invoke an explanation related to the ionization state of the infalling gas around \LAEs\ .

While this possibility relates the \LAEs\ dismissal at high redshifts to the cosmological evolution of the Universe, a different idea focuses on a parallel evolution of the \ISM\ properties.
Some encouraging hints come from observations of the line shift of the \lyman\ line deduced from other non-resonant lines, like e.g. \CII. Typically, the observed shift is $\sim 400$ km/s for \LBGs\ at intermediate redshift \citep[e.g. at $z=3$][]{Erb2014}, while \cite{Pentericci2016} find a typical shift of the \lyman\ line of only 100-200 km/s at redshift $z \approx 7$ \citep[see also][for an analysis of currently available line shifts at high redshift]{carniani:2017}.

Thus, a change in the \ISM\ properties causing a reduction of the \lyman\ line velocity shift might represent a viable solution to this puzzle, as the reduced line shift would render the \lyman\ radiation more susceptible to attenuation by the \IGM. If this is the correct explanation, one expects a positive correlation between the velocity shift of the \lyman\ line and its escape fraction.

Moreover, the connection between \LAEs\ and \LBGs\ has also remained elusive \citep{Dayal2012}. Clustering analysis at intermediate redshifts suggests an overlap of the populations, taking the form of an effective duty-cycle, with \LBGs\ for some fraction of time attaining a much larger escape fraction and turning into a \LAE~\citep{Kovac2007}. Such a duty-cycle scenario is typically quantified in terms of the fraction of dark matter halos hosting a \LAE. For example, from the SILVERRUSH survey data, \cite{Ouchi2018} infer duty cycles of $<1\%$ at $z \sim 6$  using halo occupation models, while \cite{Sobacchi2015} estimate it to be less than few percent from combining observational data with modelling of the EoR. However, the mechanism that governs this transition of a galaxy from \lyman-dark to \lyman-bright (or vice versa) remains unclear.

Theoretically, \cite{Pallottini2017} performed high-resolution zoom simulations of a prototypical high-redshift \LBG\ called \quotes{\althaea}.
By $z=6$, \althaea~has a stellar mass of $M_\star\simeq 10^{10} \msun$ and a $\rm SFR\simeq 100\,\msunyr$. It shows an exponentially rising SFR and a SFR-stellar mass relation compatible with that derived from high-$z$ observations \citep[e.g.][]{Jiang2016}. It also closely follows the Schmidt-Kennicutt relation \citep{kennicutt:2012,krumholz:2012apj}.
In \citet{Behrens2018} we studied the dust \FIR\ emission from \althaea. The resulting high dust temperatures we found might solve the puzzling low infrared excess recently deduced for the dusty galaxy A2744\_YD4 observed by \citet{Laporte2017} \citep[also see][]{Katz2019}.

In this paper, we study the \lyman\ line transfer in \althaea, focusing on observables like the \lyman\ escape fraction, line shift, and spectra along different lines of sight. We augment the original simulation with detailed \RT\ calculations of the \lyman\ radiation emitted by young stars, complementing the analysis of the UV and \FIR\ continuum \citep{Behrens2018}, and of the \FIR~emission lines \citep{Pallottini2017,vallini:2018}.
By combining these observables, we aim at understanding the relative importance of geometry, dust content, velocity field and viewing angle in determining the fraction of escaping radiation.

\section{Methods}\label{sec_methods}

First we summarize the main characteristics of the adopted hydrodynamical simulations (Sec. \ref{sec_hydro}), then we detail the assumption for the stellar emission model, dust absorption, and treatment of hydrogen ionization (Sec. \ref{sec_assumption}), and finally we describe the \RT\ simulations (Sec. \ref{sec_rt}).

\subsection{Hydrodynamical simulations}\label{sec_hydro}

We use the simulations described in \cite{Pallottini2017}. The simulations are based on a modified version of the publicly available \code{RAMSES} code \citep{teyssier:2002}. The simulations evolve a comoving cosmological volume of $(20\, {\rm Mpc/h})^{3}$, focusing on a halo of mass $10^{11} \msun$ hosting the galaxy \althaea~in a zoom-in fashion. The gas mass resolution in the zoomed region is $2\times 10^4\msun$ and the adaptive mesh refinement (AMR) allows us to reach a resolution of $\simeq 25\, \rm pc$ by $z=7$.  Outside of the zoom-in regions, our grid has a spatial resolution of 14 kpc.

In the simulation stars form from molecular hydrogen, whose abundance is computed from non-equilibrium chemistry using the \code{KROME} package\footnote{\url{https://bitbucket.org/tgrassi/krome}} \citep{grassi:2014mnras,bovino:2016aa}.
Stellar feedback is modeled as described in \citet{pallottini2017a} and includes supernova explosions, winds from massive stars, radiation pressure and accounts for the sub-grid evolution of the blastwave within molecular clouds. In the simulation stellar clusters are assumed to have a \citet{Kroupa2001} \IMF.

Most of the results presented in this work refer to the snapshot at $z=7.2$, i.e. the same one used in the fiducial model in \citet{Behrens2018}\footnote{Note, that in \citet{Behrens2018} the galaxy was shifted to a higher redshift for better comparison with observations from \citet{Laporte2017}.}. At this redshift \althaea\ has $\mathrm{SFR} \simeq 78\,\msunyr$, $M_{\star}\simeq 10^{10}\msun$, and an age of about $513\,\myr$.

\subsection{Emission and absorption properties}\label{sec_assumption}

\subsubsection{Stellar emissivity}\label{sec_stars}

\begin{figure}
\centering
\includegraphics[width=0.49\textwidth]{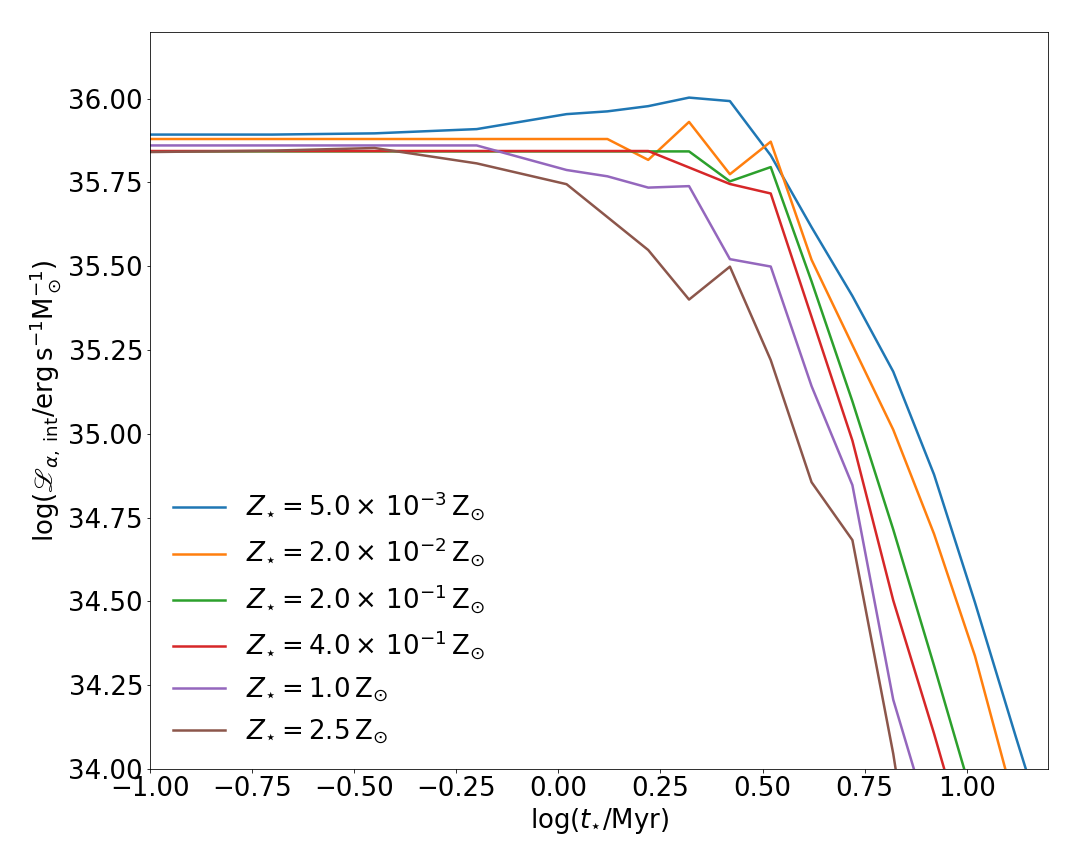}
\caption{Intrinsic \lyman\ emission ($\mathscr{L}_{\alpha,\rm int}$) from stellar sources per stellar mass as a function of stellar age ($t_{\star}$); different lines indicate different stellar metallicities ($Z_{\star}$).
\label{fig_intrinsic}
}
\end{figure}

In \althaea\ we keep track of age ($t_\star$) and metallicity ($Z_{\star}$) of all its stellar clusters. Therefore, we can use population synthesis codes to derive the \SED\ of each stellar cluster. In particular, we use \citet{Bruzual2003} prescribing the same \citet{Kroupa2001} \IMF\ assumed for the feedback in the hydrodynamical simulations.

These assumptions yield the intrinsic continuum flux directly. Integrating the ionizing part of each \SED, we obtain the total ionizing power $L_{\mathrm{ion}}$. Assuming Case-B recombination and a gas temperature of 10$^4$ K, the latter can be related to the intrinsic \lyman\ radiation via \citep{Dijkstra2014}
\begin{equation}
L_{\alpha} = 0.68 L_{\mathrm{ion}}(1-f_{esc})\,.
\end{equation}
We will assume in the following that the escape fraction of ionizing photons, $f_{esc}=0$. This maximizes the \lyman\ output. We note that the factor of 0.68 is only strictly valid at a temperature of $10^4$ K. However, as the temperature dependence is very weak (at 10$^3$ K it is 0.77), we neglect it.

Practically, we use a table and interpolate age and metallicity to obtain the rate of ionizing photons. In Fig. \ref{fig_intrinsic}, we show this rate for different $Z_{\star}$ as a function of $t_{\star}$. We assume that the stars in a cluster form in a single burst, and treat the stellar clusters as point emitters of \lyman\ radiation.

We do not consider any additional source of \lyman\ radiation. In particular, we do not include cooling radiation from hydrogen around temperatures of $\sim 10^4$ K. The uncertainties in estimating the intrinsic luminosity of \lyman\ in cooling radiation are large \citep[e.g., see][]{Goerdt2010,Rosdahl2012,FaucherGiguere2010}. As an upper limit, we can calculate the intrinsic \lyman\ luminosity assuming that all the gas with temperature $10^4<T/{\rm K}<5\times10^4$ cools only due to \lyman\ radiation. In this case, we find that the contribution from cooling radiation is 1\% of the intrinsic luminosity coming from stellar ionizing flux. Hence we have decided to neglect this contribution.

\subsubsection{Dust model}\label{sec_dust}

The evolution of the gas and stellar metallicity is traced in \althaea. However, we stress that the simulation does not keep track of dust formation and/or destruction processes, and therefore does not follow the evolution of dust content/properties directly. 
Therefore, we need to make an assumption for the conversion between metallicity and dust content. We take the same route as in \cite{Behrens2018} by assuming a linear relation between metal and dust mass, with a constant conversion factor $f_\mathrm{d}=0.08$, and use the dust model developed by \cite{Weingartner2001,Li2001} for the Milky Way (MW). Our value for $f_\mathrm{d}$ is lower than the one determined for the MW ($\sim$ 0.3-0.5). The choice is motivated by the constraints obtained by \cite{Behrens2018} in order to fit the ALMA observations from \cite{Laporte2017}. With this value, the dust mass in \althaea~is about $1.6\times 10^{6}\msun$. As we will see, the exact value does not affect our results substantially.

We note that \citet{Laursen2009} have developed a model explicitly taking into account that dust may have a different abundance in ionized regions (like \HII\ regions). This translates into reducing $f_\mathrm{d}$ in such regions in our notation. We do not follow this route, but will analyze an extreme case (setting $f_\mathrm{d}$ to zero in \HII\ regions), see the next section.
\subsubsection{Hydrogen ionization}
In order to do the \lyman\ RT, we need to estimate the neutral hydrogen content, i.e. the fraction of hydrogen in ionized and molecular state must be calculated.
The hydrodynamical simulation provides the molecular hydrogen fraction of the gas. To obtain the ionized fraction of gas, we assume the prescription of \cite{Chardin2017}, which extends and refines the model of \cite{Rahmati2013} for $z>6$.
Such prescription accounts for collisional excitation and a \citet{haardt:2012} UV background radiation, that is suppressed at high gas densities. At typical \ISM\ densities, the UV background plays essentially no role \citep[e.g.][]{gnedin:2010}, as on these scales local sources are the dominant producers of ionizing radiation.

To recover the effect of local sources, we post-process the hydrodynamical simulation by considering \HII\ regions around young stars. The size of an \HII\ region can be estimated by the radius of a Str\"omgren sphere: 
\begin{equation}
R_S^3 = \frac{3Q}{4 \pi n^2 \alpha_B}
\end{equation}
with $Q$ being the ionizing photon rate, $n$ the gas number density, and the Case-B recombination rate $\alpha_B = 2.6\times 10^{-13}$ cm$^3$/s.
Consider that stellar clusters in the simulation have masses of about $2 \times 10^4$ $\msun$; with the chosen \IMF\ and stellar \SED\ (see Sec. \ref{sec_stars}), we find $Q \sim 4 \times 10^{50}$ s$^{-1}$, in turn yielding $R_S \sim 10$ pc for $n=10^{2} {\rm cm}^{-3}$. Such density is selected since it is the typical of star forming regions in the simulation and it is consistent with the expected one in \HII\ regions \citep[e.g.][]{Dopita1986,Kewley2002,Kewley2013}. Note that the Str\"omgren solution is reached on timescales of $\sim 10^4$ yr while pressure equilibrium is marginally achieved on time scales of 10$^6$ yr, when photoevaporation effects start to play a role \citep{decataldo:2017}, whereas the stars responsible of the ionizing flux have lifetime of around 10$^7$ yr \citep[e.g. see][]{Lejeune2001}.

This simple estimate neglects the complex evolution of \HII\ regions, the history of each star-forming region, the interaction of overlapping region, and the effect on gas dynamics. In our simulations, the linear size of the most-refined cells is 25 pc, and we therefore chose to treat each of the cells hosting stellar cluster as ionized bubble, setting the gas neutral fraction to zero\footnote{If a cell is not refined to the finest level $L_{max}$, we set its partial ionization by ionizing a physical volume of (25 pc)$^3$ in the cell, that is, we set the ionization level to 1/$(L_{max}-L)^3$, where $L$ is the refinement level of the cell in question.}. 

The adopted method gives a crude approximation of the radial profile within \HII\ regions, however it is justified since we do not have the spatial resolution to resolve the ionized bubbles.
Setting the neutral fraction to zero in regions around young stars maximize their effect on the \lyman\ RT, since the gas becomes transparent to the \lyman\ radiation. Because of this, we do not need to modify the temperature or velocity fields in the \HII\ regions. In total, our prescription turns about $7 \times 10^8 \msun$ of neutral gas into ionized gas, i.e. about 20\% of the gas mass in the ISM.

As stated above, this prescription is part of our fiducial setup; below, we will sometimes compare to the results we obtain when we instead completely ignore local sources of ionizing photons. We will refer to this model as \quotes{no \HII\ bubbles} model.
\subsection{Radiative transfer}\label{sec_rt}

For the \lyman\ \RT, we use our own code, \code{Iltis} (Sec. \ref{sec_iltis}). 
For the continuum UV \RT\ (Sec. \ref{sec_dust_rt}), we make use of the public \code{Skirt}\footnote{version 8, \url{http://www.skirt.ugent.be}} code \citep{Baes2003,Baes2015,Camps2015,Camps2016}.
While these calculations rely on Monte-Carlo methods, the \CII\ emission can be modeled using an semi-analytic approach (Sec. \ref{sec_cii_model}). We describe the three approaches in more detail below.

\subsubsection{\lya line}\label{sec_iltis}

We use an updated version of the code \code{Iltis} already used in \citet{Behrens2013,Behrens2014a,Behrens2014b,Behrens2018a}. We refer the reader to these papers for more details on the implementation, and to the public code repository\footnote{\url{https://github.com/cbehren/Iltis}}. In the following, we briefly summarize the \RT\ algorithm. 

\code{Iltis} makes use of the usual Monte-Carlo approach for \lyman\ \RT; it follows the escape of a significant number of tracer photons from their emission sites through the simulation volume\footnote{Note that \code{Iltis} can directly handle Ramses datasets, using a modified version of the \code{RamsesRead++} library originally written by Oliver Hahn (see \url{https://bitbucket.org/ohahn/ramsesread}).}. Photons can be scattered by gas, or scattered/absorbed by dust. In case they are absorbed by dust, they are considered \quotes{lost}, since reemission will take place in the IR regime.
The optical depth a photon experiences is determined by the density of the neutral gas along its path, the gas temperature, the bulk velocity of the gas, the density of dust, and the photon frequency.
Initial frequencies, points of interaction, thermal velocities of scattering atoms, and scattering angles are drawn from appropriate probability distributions. On top, we also consider the redshifting of photons due to the Hubble flow, whose effect is to shift photons at wavelengths shorter than the line center (\quotes{blue} photons) back into the line center, rendering them subject to a large attenuation even in the diffuse IGM, far away from the ISM.

In order to produce meaningful surface brightness maps, we use the peeling-off technique \citep[e.g.][]{Zheng2002,Dupree2004}. At each scattering event, we evaluate the probability that the photon directly escapes into the direction of the observer, which can be understood as weighting this contribution with the total flux escaping in the specified direction. This probability can be written as
\begin{equation}
P = p(\theta)\exp(-\tau_o)\,
\end{equation}
where $p(\theta)$ is the probability to be scattered in the direction of the observer, and $\tau_o$ is the optical depth the photon would penetrate before reaching the observer.
Technically, this means that we need to integrate the optical depth from the point of interaction throughout the simulation box.

We perform radiative transfer at the spatial resolution of the hydro simulation (25 pc). For comparison, \cite{Smith2015} performed \lyman\ RT at considerably higher resolution (up to 830 au or 0.004 pc), but in a smaller box, as they considered halos of mass $\sim$ 10$^7 \msun$. \cite{Trebitsch2016} reached 434 pc; \cite[][also see their table 1 for more references]{Verhamme2012} and \cite{Behrens2014a} reached 18/30 pc, but with an idealized setup for an isolated disk galaxy and no cosmological initial conditions. Recently, \cite{Smith2019} presented high-resolution simulations ($<10 pc$) for a \LAE\ with a stellar mass of about $5 \times 10^8 \msun$.

 More technical details about the numerical setup can be found in App. \ref{sec_app_rt}. In brief, we launch $10^{3-4}$ photon packages per source from the center of the hydro cell. The intrinsic spectrum is a Gaussian of width 10 km/s, and we use a standard acceleration scheme to avoid core scatterings \citep{Dijkstra2006}.

When presenting the results, note that our definition of the \lyman\ escape fraction $f_\alpha$ is related to the flux actually observable along an individual line of sight (los). It therefore combines two very different mechanisms: (a) direct loss of photons due to absorption by dust in ISM on the one hand, and (b) indirect losses due to diffuse scatterings out of the los in the CGM/IGM. In the former case, we expect the photons to be re-emitted in the IR; in the latter, we expect them to contribute to a diffuse background or an extended halo. On top, we do consider here the escape along one specific los; even in a dust-free galaxy, and without considering the IGM, we expect anisotropic escape to take place due to the relative distribution of gas/stars and peculiar velocity fields.

To recover observables like the line flux from the galaxy, the escape fraction, or spectra, we integrate over a circular aperture of radius 2 kpc around the center of the galaxy. We chose this size of the aperture to include only the escaping radiation from \althaea, which has a size of about 1 kpc, and not the diffuse background coming from \lyman\ scattered in the CGM/IGM.

\subsubsection{UV and IR continuum}\label{sec_dust_rt}

We make use of \code{Skirt}, a publicly available code for detailed simulations of UV continuum, absorption by dust, and re-emission of the absorbed energy in the IR. The code is very flexible thanks to its ability to load e.g. data from AMR or Smoothed-particle hydrodynamics (SPH) simulations, a variety of implemented, \SED\ (Sec. \ref{sec_stars}), dust models (Sec. \ref{sec_dust}), and physical mechanisms.

Since we are mostly interested here in the escaping UV continuum rather than the mid/\FIR\ emission, we do not take into account self-absorption of dust or stochastic heating. Note that the modeling of the UV/IR emission used here is the same as in \citet{Behrens2018}.

\subsubsection{ \CII~line emission}\label{sec_cii_model}

We compute \CII~with a semi-analytical method detailed in \citet[][see appendix C]{pallottini2017a}. The basic features are as follows.
First we compute the column density of \CIIion~in each cell by using a grid of photoionization models obtained with  \code{cloudy}\footnote{version 13.03, see \url{https://www.nublado.org/}} \citep{cloudy:2013}. For the \code{cloudy} calculations, we assume a uniform UV interstellar radiation field with a MW-like spectral shape and an intensity scaled to the Althaea SFR. Then, the \CII~emissivity is computed as a function of gas temperature and density following \citet{dalgarno:1972}, and  \citet{vallini:2013,vallini:2015}. 
Finally, we account for CMB flux suppression as in \citet[][]{vallini:2015,pallottini:15}. As the [CII] line is observed against the CMB, the spin temperature of the transition must differ from the CMB one to be detectable. As this can be obtained essentially only via collision with other species, the emission for low density gas is largely suppressed. This is an important effect that needs to be included. 

In the present paper the \CII\ emission is used only to calculate its velocity shift with respect to the \lyman\ line. A full analysis of the kinematic/morphology of the \CII\ line from \althaea\ is presented in \citet{kohandel:2018}. In order to acquire meaningful line shifts, in this work the systemic velocity of the galaxy is taken to be the one of the \CII\, which is determined by fitting the \CII~spectrum of \althaea\ for each los with a Gaussian.

We note that \CII\ is optically thin, and therefore solely affected by the local conditions of the emitting gas.

\section{Results}\label{sec_results}

\subsection{Morphology and \lyman\ escape fraction}

\begin{figure*}
  \centering
  \includegraphics[width=0.8\textwidth]{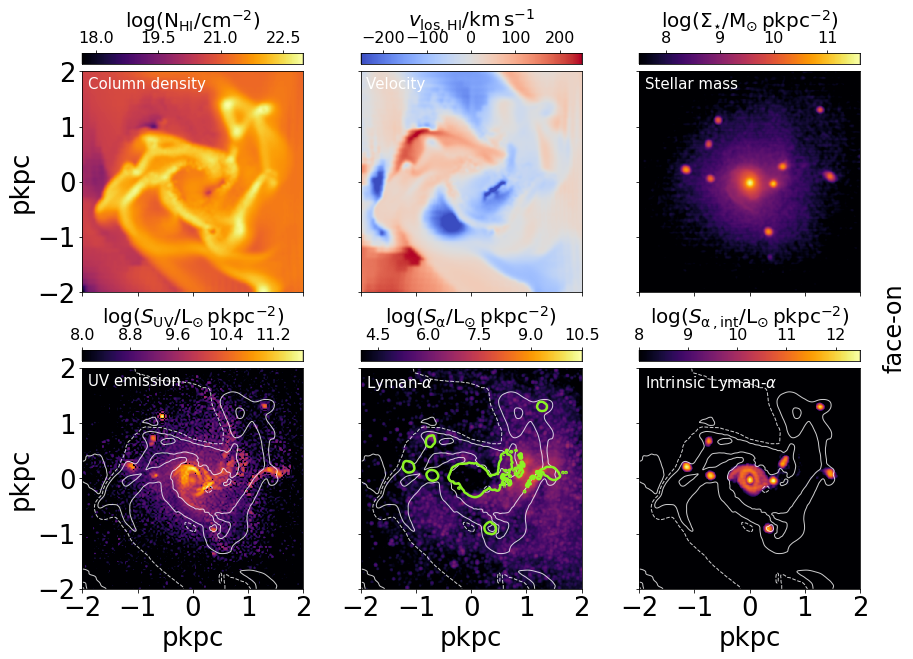}
  \includegraphics[width=0.8\textwidth]{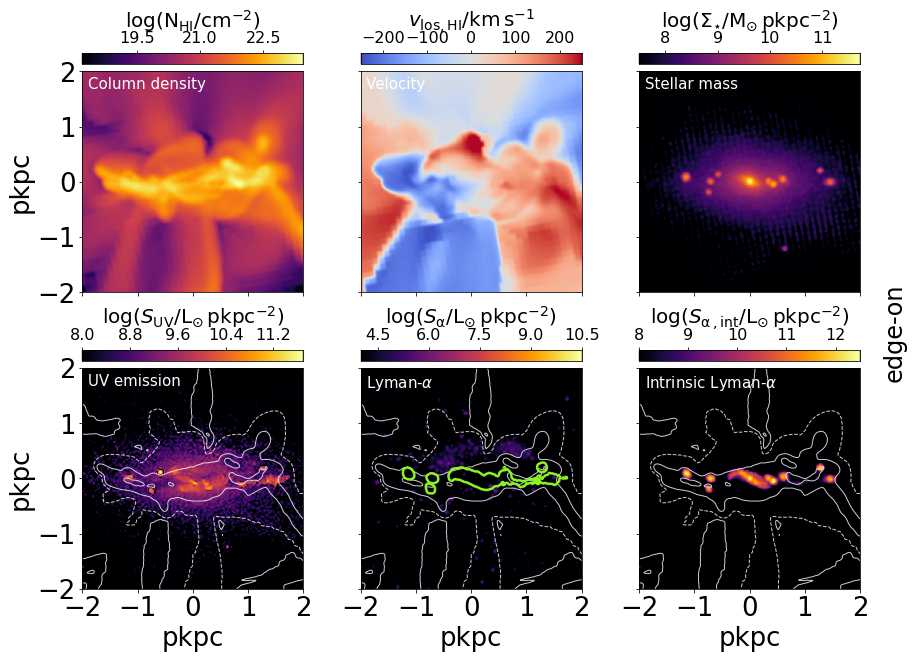}
  \caption{Morphology of \althaea\ for two different lines of sight, face-on (top) and edge-on (bottom).
For each los, in the upper panels we show the neutral hydrogen column density ($\rm N_{\rm HI}$, left), the los velocity ($v_{\rm los,HI}$, center) and stellar surface density ($\Sigma_{\star}$, right), while in the lower panels we show the UV surface brightness ($S_{\rm UV}$, left), \lyman\ surface brightness ($S_{\alpha}$, center
), and the intrinsic \lyman\ surface brightness ($S_{\alpha,\rm int}$, right).
In the bottom panels, dashed, thin (white) contours correspond to a hydrogen column density of $10^{22}$ cm$^{-2}$. Inner (outer) contours denotes higher (lower) column densities, spaced by 1 dex. The thick (green) contours in the bottom center panel show the location of the intrinsic \lyman\ emission. Note that the velocity map shows the mass-weighted velocity integrated along the los.
\label{fig_morphology}}
\end{figure*}

We start by analyzing the galaxy morphology resulting from our fiducial model ($z=7.2$). This is shown in Fig. \ref{fig_morphology} for two lines of sight, chosen to be face-on and edge-on, respectively.
We first focus on the face-on case. As \althaea\ is very compact, the mean \textit{neutral} hydrogen column density is $N_{HI} = 6 \times 10^{21} \mathrm{cm}^{-2}$; in the proximity of the star-forming clumps, the typical neutral column density is larger ($\sim 3 \times 10^{22} \mathrm{cm}^{-2}$); however, since the plot takes into account the ionized bubble model outlined above, some of the clumps display a core of reduced column density, owing to the ionization from local sources.

The los component of the velocity of the neutral gas, $v_{\rm los,HI}$, shows a very complex pattern, owing to the dynamics of accretion through filaments, tidal stripping of gas streams from nearby satellites, and SN-driven outflows. The distribution of stellar mass has a half-mass radius of 0.5 kpc and shows a diffuse stellar component extending to about 1 kpc, also featuring multiple clumps in the disk. Typical velocities within the disk are few $\sim 100$ km/s.
The diffuse component mainly consists of relatively old stars, as star formation in \althaea\ takes place mostly in the central region and in the (relatively small) clumps \citep[see][]{Behrens2018}. Such star forming clumps are characterized by high stellar mass surface densities ($\Sigma_{\star}\sim 10^{11}\surfd$). 
Young stars producing most of the intrinsic \lyman\ radiation are predominantly found in these regions, as clearly shown by the corresponding map of the intrinsic \lyman\ emission $S_{\alpha,\rm int}$. Clumps can reach surface brightness of $S_{\alpha,\rm int}\sim 10^{12}\surfl$.

By comparing $S_{\alpha,\rm int}$ with the surface brightness of the escaping \lyman\ emission, $S_{\alpha}$, one sees that \lyman\ radiation escapes virtually only from the south-west side of the disk, with a peak located in an inter-arm region at about 1 kpc from the center. Such region is characterized by a low gas column density ($N_H\sim 10^{19}\cold$), and shows the signature of an outflow expanding at a velocity $v_{\rm los}\sim 100$ km/s along the los.

In general the UV emission is more diffuse than the \lyman\ one. While the central region is Ly$\alpha$-dark, it is relatively bright in UV; the same is true for most of the star-forming clumps. However, the UV continuum map also shows a surface brightness peak co-located with the \lyman\ peak. Both observations can be explained as a consequence of the larger attenuation of the resonantly scattered \lyman\ radiation: In regions that allow \lyman\ to escape, UV will escape as well.

The galaxy-averaged \lyman\ escape fraction is $f_\alpha = 3 \times 10^{-4}$; while the intrinsic \lyman\ luminosity of \althaea\  is $2 \times 10^{44}$ erg/s, only $5 \times 10^{40}$ erg/s reaches the observer\footnote{We have checked that the global morphology and escape fraction do not vary appreciably if we drop the fiducial model assumption that the gas around stars is ionized.}. Recall that in the computation of $f_\alpha$ we include the effects of both the ISM and CGM/IGM (see Sec. \ref{sec_iltis}). For comparison, the UV escape fraction (evaluated at 1500 \AA{}) in this case is $f_{\mathrm{UV}}=0.05$. 

For the fiducial case, given the FWHM of $\sim$ 2 \AA{}, we obtain a total Ly$\alpha$ flux of $\sim 10^{-19}$~erg~s$^{-1}$cm$^{-2}$. The detection of these Ly$\alpha$ fluxes is challenging for current telescopes \citep[but see][]{Wisotzki2018}, but still achievable with future instruments. For example, the Multi-Object Spectrograph planned for E-ELT, is expected to achieve a sensitivity of $\sim 10^{-19}$erg~s$^{-1}$cm$^{-2}$ in 20 hr of observing time \citep{evans2015}.

We show the same set of maps also for the edge-on direction\footnote{The orientation of the plots has been chosen such that the face-on direction discussed before is pointing upwards (along the y axis) in these panels.} in Fig. \ref{fig_morphology}.
The column densities in this case are obviously higher. We can see that while the disk is relatively thin, it is far from being flat, owing to the complex gas motions resulting in a bent and warped disk structure. This is not only visible in the gas density and kinematical structure, but also in the distribution of stars. 

As expected, the escape of \lyman\ radiation is much more difficult in this direction: in fact, we find  $f_\alpha = 5 \times 10^{-6}$, i.e. about 60 times lower than in the face-on case. The \lyman\ transmission in the edge-on case exclusively comes from photons that are scattered into this los at several 100 pc above the disk. We also note a strong above/below asymmetry, that is, photons preferentially come from above the disk. This asymmetry is confirmed by the more systematical variation of the inclination angle discussed in Sec. \ref{sec_tess}. For comparison, $f_{\mathrm{UV}}=0.02$ in this case. 

Finally, we derive the \lyman\ line EW.  While the intrinsic value obtained by considering the attenuated \SED\ of the stars in \althaea\ is 103 \AA{}, the \EW\ after \RT\ decreases to 0.7 (0.04) \AA{} for the face-on (edge-on) case.

\subsection{Inclination effects}\label{sec_tess}

\begin{figure}
\centering
\includegraphics[width=0.53\textwidth]{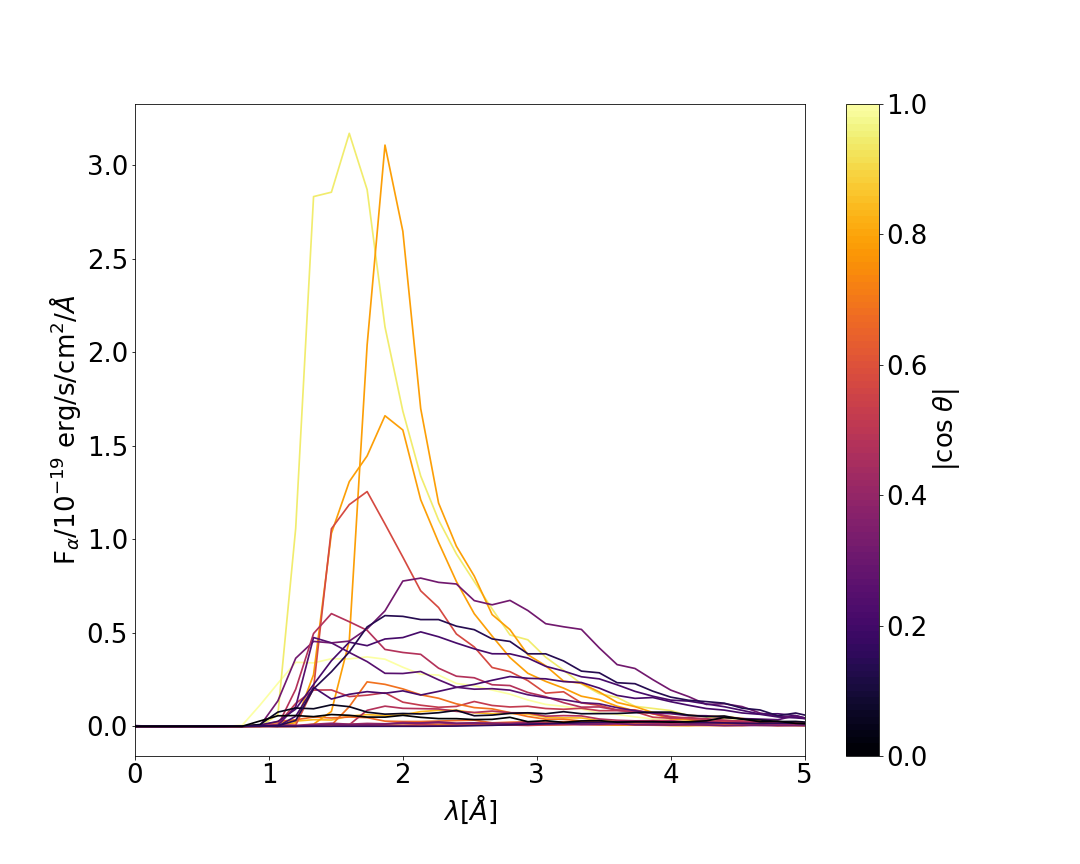}
\caption{\lyman\ spectra from the fiducial simulation along various lines of sight, shown as spectral flux F$_\alpha$ versus wavelength. The colors indicate the inclination of the corresponding los, ranging from 0 (edge-on) to 1 (face-on).
\label{fig_tess_spectra}
}
\end{figure}  

\begin{figure*}
\centering
\includegraphics[width=0.32\textwidth]{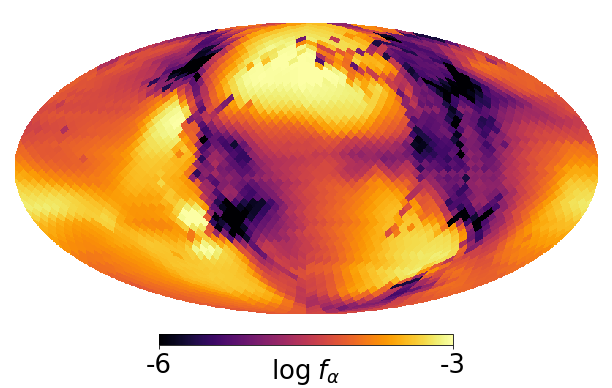}
\includegraphics[width=0.32\textwidth]{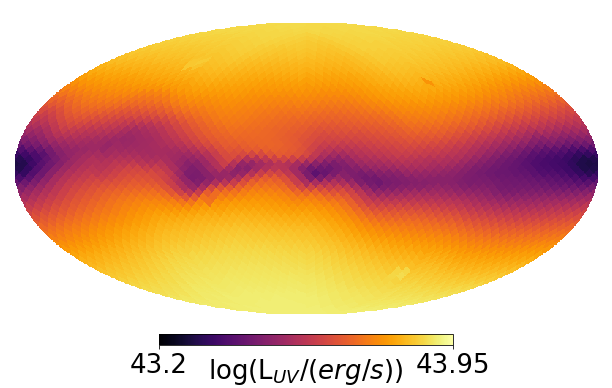}
\includegraphics[width=0.32\textwidth]{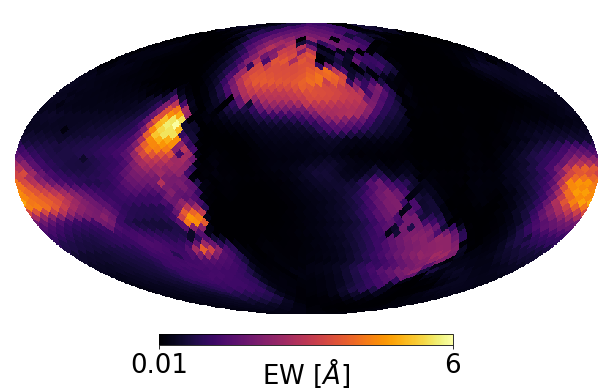}
\caption{Mollweide plots of the \lyman\/UV after the RT. In the left panel we show the escape fraction ($f_{\alpha}$), the UV luminosity in the center panel, and in the right panel we show the EW. Face-on direction corresponds to the top/bottom center.
\label{fig_tess_mollweide_lymanew}
}
\end{figure*}  

Several authors \citep{Laursen2009, Verhamme2012, Behrens2014a, Behrens2014b} have pointed out that due to its resonant nature, the \lyman\ line is particularly susceptible to geometrical effects or viewing angles, owing to the large optical depth \lyman\ photons experienced from neutral gas compared to UV continuum photons. However, this conclusion is geometry-dependent. If photons travel in a diffuse \ISM\ in which dense, optically-thick clouds are embedded, it can be shown that under certain conditions \lyman\ the escape of \lyman\ photons is enhanced as they simply bounce off the surface of the dense clouds instead of penetrating them like continuum UV photons. However, this so-called Neufeld effect \citep{Neufeld1991} is only valid in a very narrow regime of parameters for e.g. the velocity dispersion of the clouds \citep{Laursen2013,Duval2014}. Hence, in general, it is expected that the inclination effect leaves a stronger imprint on the \lyman\ properties. We investigate the possible inclination effect on our results in the following.

To this aim, we have generated 3072 los by using an equal area and iso-latitude tessellation \citep{Healpy}, and ran \lyman~and UV/IR simulations for all of them.
For illustration, the \lyman\ spectra of some los are shown in Fig. \ref{fig_tess_spectra}. All the spectra show a red peak as expected, but the total line luminosity, and position of the peaks vary from los to los.

\begin{figure}
\centering
\includegraphics[width=0.49\textwidth]{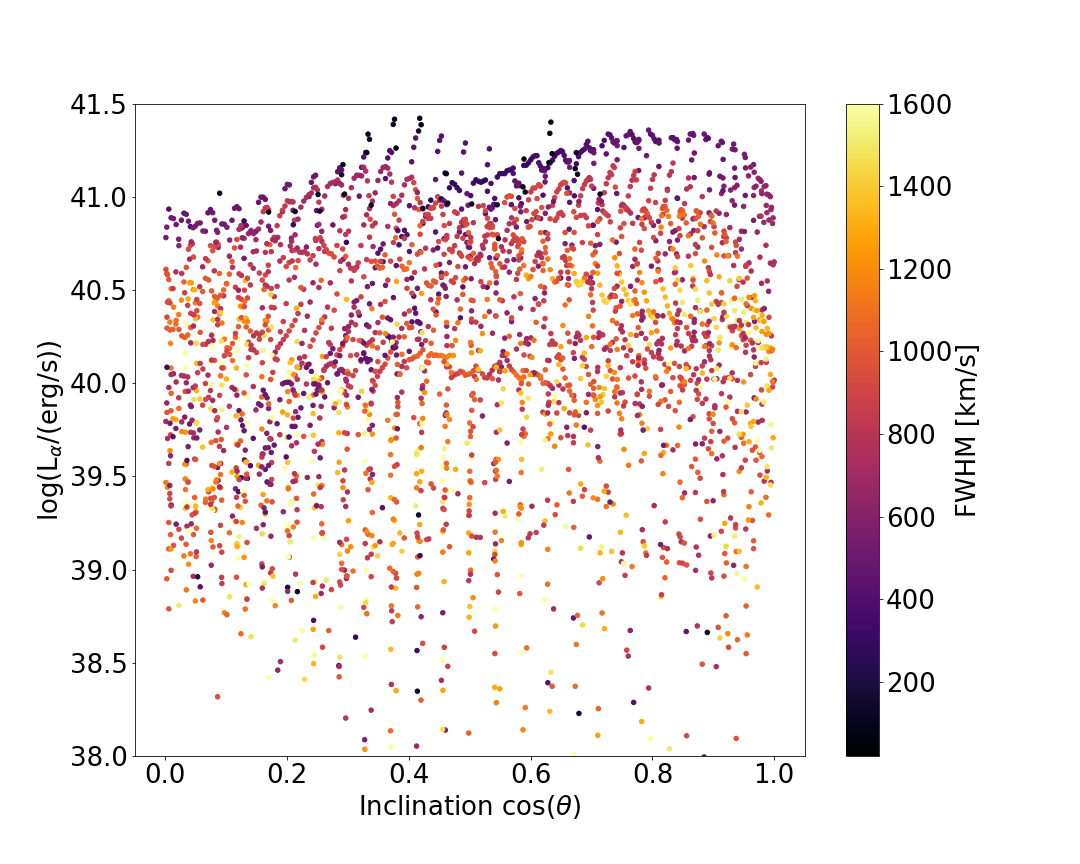}
\caption{\lyman\ line luminosity as a function of inclination. Colors indicate the corresponding line width (FWHM).\label{fig_linewidths}}
\end{figure}

In the left panel of Fig. \ref{fig_tess_mollweide_lymanew}, we show the escape fraction from each los in a Mollweide plot. $f_\alpha$ varies by about six orders of magnitude and the brightest los reaches a luminosity of about $5 \times 10^{41}$ erg/s. The line width (FWHM) varies largely from 60 km/s to 2000 km/s.Observationally, line widths of few hundred km/s are typical at both lower redshift \cite[e.g. see][figure 2] {Verhamme2018} and higher redshift \citep[e.g.][]{Vanzella2011,Matthee2017,Pentericci2018}.
In Fig. \ref{fig_linewidths}, we show the \lyman\ luminosity, inclination, and line full-width half-maximum (FWHM) for each of the 3072 lines of sight considered. A value $\cos\theta=0$ ($\cos\theta=1$) corresponds to a edge-on (face-on) orientation. In contrast to e.g. \cite{Verhamme2012,Behrens2014b}, the preferential \lyman\ escape along face-on los is less pronounced; the behavior is instead much more random. This can be understood by considering that  \cite{Behrens2014b, Verhamme2012} used an idealized, non-cosmological setup for a disk galaxy, whereas \althaea\ is a complex object in its cosmological environment, with gas flowing in along filaments, a warped disk with many small-scale features, and satellites swirling around the main disk. Comparing with the results of \cite{Laursen2013}, we do not find a significant increase of the \EW\ in face-on directions, similar to what \cite{Yajima2012} found in a simulated galaxy at intermediate redshift ($z=3.1$). In the latter case,  inclination effects become prominent only at later times ($z\approx 0$), when the galaxy has settled into an ordered disk.

Consistent with the results in the above Section, the dependency of the UV continuum on the viewing angle (Fig. \ref{fig_tess_mollweide_lymanew}, center) is shallower compared to the \lyman; the emerging UV luminosity varies with inclination by a factor of 4 at most \citep[see also Fig. B1 in][]{Behrens2018} and it is essentially invariant with respect to the azimuthal angle. The \lyman\ line EW for every individual los is shown in the right panel of Fig. \ref{fig_tess_mollweide_lymanew}. Its angular distribution broadly correlates with the $f_\alpha$ map, and shows a maximum value of 5.9 \AA{}. 

\subsection{What quenches the \lyman\ line?}\label{sec_explanation}

\begin{figure}
\centering
\includegraphics[width=0.49\textwidth]{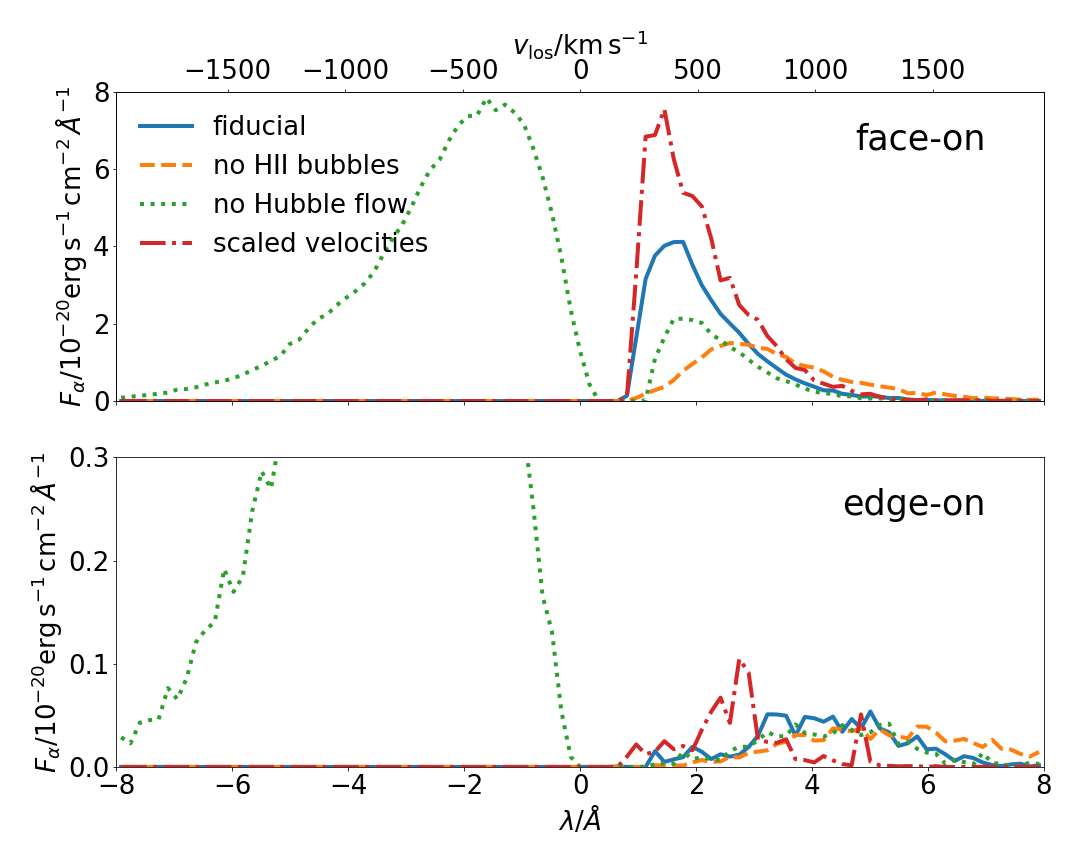}
\caption{Spectra obtained from \althaea, along two different lines of sight: face-on (top) and edge-on (bottom). Additionally to the fiducial model, we also show the spectra obtained by ignoring local \HII\ regions (no \HII\ bubbles), the spectra obtained by switching off the Hubble flow (no Hubble flow), and the spectra obtained by reducing the velocity field magnitude (scaled velocities).\label{fig_spectra}}
\end{figure}
\begin{figure}
\centering
\includegraphics[width=0.49\textwidth]{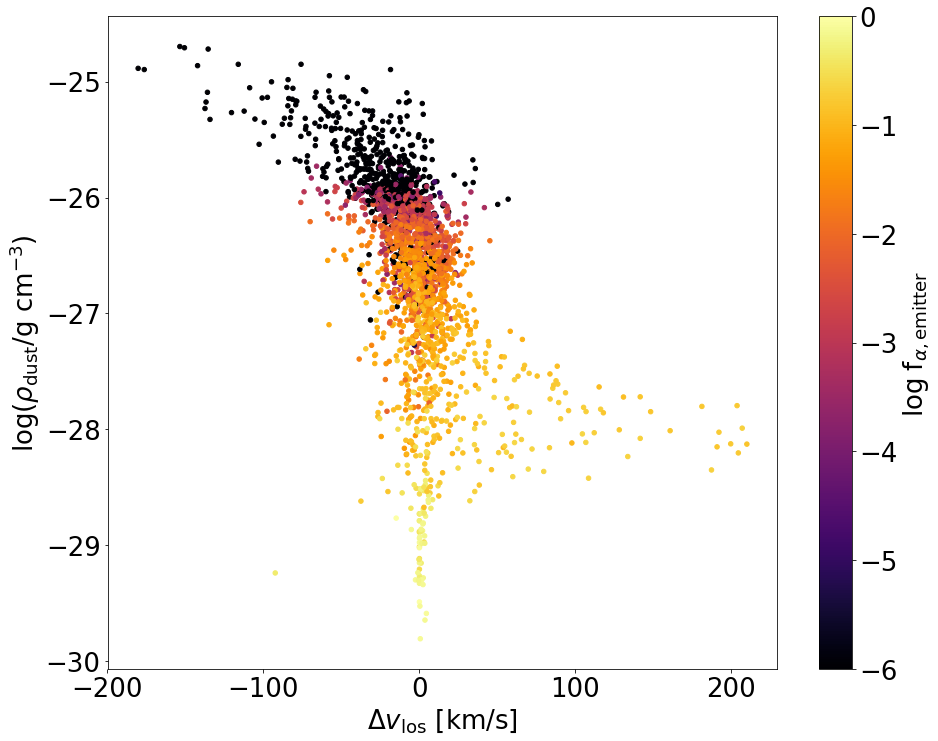}
\caption{Correlation between the \lyman\ escape fraction $f_{\alpha,\mathrm{emitter}}$ for all \textit{individual} stellar clusters (values given by the colorbar), dust mass density at the location of the emitter, and velocity differential along the los; positive values indicate outflow kinematics.}.
\label{fig_vel_dust}
\end{figure}

To get a deeper insight on the physics determining the low $f_\alpha$ values found, it is useful to inspect the line spectra along different los. 
We build spectra for the two los shown in Fig. \ref{fig_morphology}.

In Fig. \ref{fig_spectra}, we show the resulting \lyman\ spectra for the face-on (top) and edge-on (bottom) cases. Both the fiducial run and the \quotes{no \HII\ bubble} case are shown, together with some tests discussed below. In the following, we try to disentangle different effects determining the resulting escape fraction and the shape of the spectrum.

\subsubsection*{IGM attenuation}
The blue wing of the line is almost completely blanketed by the intervening IGM. This is a well known effect \citep{Dijkstra2006,Laursen2010} caused by the Hubble flow: photons leaving the galaxy blue-wards of the \lyman\ line center will redshift into the line center where they are scattered out of the los. Indeed of the \lyman\ radiation escaping the ISM is blue-shifted, indicating that infall (rather than outflow) dominates the overall dynamics. This is not surprising, as \althaea\ is by far the most massive object in its environment.
 
IGM quenching can be distilled by switching off Hubble expansion in the simulation. We stress that we do not render the surrounding gas outside of the galaxy, but only switch off the Hubble expansion. In this case we find a \quotes{classical} double-peaked spectrum with a flux of $2 \times 10^{41}$ erg/s in the blue peak (dotted line in Fig. \ref{fig_spectra}). This blue peak is completely erased by the IGM in the presence of the Hubble flow, while the red peak is not strongly affected (given that at $z\approx 7.2$ the mean IGM ionization is $x_{\rm HI} \approx 10^{-3}$ in the simulation). Note that switching off the Hubble flow changes the velocity field that photons experience everywhere; this is the reason why the red part of the spectrum changes when the Hubble flow is turned off. In the case of this particular los, the effect on the red wing is quite strong, probably owing to the specific density structure above the disk along this los; for other los, it is at the 10\% level, so this has to be seen as a peculiarity of the los.

We can factorize the total escape fraction as $f_\alpha = f_{\mathrm{ISM}} f_{\mathrm{IGM}}$, i.e. the fraction of photons escaping the galaxy times the fraction of photons reaching us through the IGM. For the face-on case, the fiducial model predicts  $f_{\mathrm{ISM}} = 0.001$, $f_{\mathrm{IGM}} = 0.23$.
If we instead of switching off the Hubble flow remove all the gas outside a radius of 30 pkpc to assess the influence of the IGM, $f_{\mathrm{IGM}}$ goes up by a factor of 1.8, and $f_{\mathrm{ISM}}$ goes down accordingly.

We conclude that while the IGM does reduce the escaping flux by a factor $\le 5$ it is not the major cause of line quenching.

\subsubsection*{Neutral hydrogen in the ISM}
The presence of neutral gas increases the scattering rate and ultimately the probability for the \lyman\ photon to be absorbed by dust. To check its importance in determining the escape fraction, in Fig. \ref{fig_spectra} we also plot the spectra from the model lacking \HII\ regions around stellar clusters (dashed lines).
These spectra deviate moderately from the fiducial ones. They exhibit a larger line shift ($\sim 1.8 $ \AA{} vs. $\sim 1.0 $ \AA{}), and the flux is reduced by a factor of $\sim$ 2.\footnote{Line shifts are obtained by fitting a skewed Gaussian to the \lyman\ line profile.}. However, in terms of the escape fraction, both the fiducial and the no bubble case differ only by factor of $\sim 2$ . In other words, increasing the HI fraction close to stars does not affect the escape fraction significantly. 
Stated differently, the gas close to the star forming sites alone does not appreciably increase the path length of \lyman\ photons, which is mainly set by the residual gas in the galaxy. 

\subsubsection*{Infall/outflows}

The ratio of the fluxes bluewards/redwards of the line center depends on the peculiar velocity field. Relatively mild outflows, with mean velocities around $150\,\kms$, are present in the galaxy; infalling streams also share the same velocity scale \citep[][]{gallerani:2018}. Infall motions are dominating the radiative transfer, as can be seen by the large amplitude of the blue peak in the spectrum when we ignore the Hubble flow (Fig. \ref{fig_spectra}, dotted line). In order to check the effect of the peculiar velocities, we re-run the \lyman\ \RT\ by down-scaling their norm by a factor of 4. 

The resulting spectrum is shown in Fig. \ref{fig_spectra} (dashed-dotted lines). As now less photons are transferred to the blue side of the line center (and more are transferred to the red side), less flux is scattered out of the los by the CGM/IGM Although $f_\alpha$ and EW are enhanced by a factor of $\simeq 2$ with respect to the fiducial model, such increase is not  sufficient to promote \althaea\ into a LAE. 

In Fig. \ref{fig_vel_dust}, we show $f_{\alpha,\mathrm{emitter}}$ for each individual stellar cluster in the fiducial simulation along the brightest los as a function of the dust density at the location of the cluster, and the velocity differential, $\Delta v_{los}$, along the los measured 25 pc above the source. Positive values of $\Delta v_{los}$ correspond to outflows. In addition to the strong negative correlation between dust density and escape fraction, there is a clear,  increasing trend of $f_\alpha$  with $\Delta v_{los}$. The effect imprinted by peculiar velocities is nevertheless sub-dominant with respect to dust absorption, as we discuss further below. 

\subsubsection*{Absorption by dust}

The vast majority of \lyman\ photons in the simulation are absorbed on the spot, that is, in the gas element they are spawned in. About 60\% of all absorptions occur in the production cell in the fiducial model, i.e. within $\sim 15$ pc from their emission spot. 
In the \quotes{no bubble model}, this fraction increases to 95\% due to the increased absorption probability driven by an increased pathlength due to more scatterings on hydrogen.
It is therefore an interesting question to ask whether the attenuation of the \lyman\ is mainly affected by (a) the total amount of dust, or (b) by its distribution, given the fact that dust is found to be highly concentrated in the star-forming knots discussed above.

To answer this question, we first down-scaled the dust mass in the whole simulation volume, that is, we reduced $f_{\mathrm{d}}$ by a factor of 10. As a result, we find that $f_\alpha$ increases from $3 \times 10^{-4}$ to $4 \times 10^{-3}$ in the face-on case. However, the UV continuum raises as well, with the net result that the EW is lower ($EW=0.3\;\AA{}$ for the face-on case).

Next, we ran a different set of simulations in which we selectively removed the dust from each star-hosting cell, that is, we rendered \HII\ regions dust-free. We refer to this model as the \quotes{dust-free bubbles} model. Since the dust is highly concentrated in the star forming knots, this procedure removes about 60\% of the total amount of dust from \althaea. Note that the total amount of dust is still 4$\times$ higher than in the previous case in which we decreased the dust content everywhere.
In this case, the escape fraction increases by a factor of 100, yielding  $f_\alpha = 0.03$ (face-on); this corresponds to a \lyman\ luminosity of $6 \times 10^{42}$ erg/s. However, as the UV increases as well, the boost of the EW is more modest (4.2 \AA{}).

This finding is also backed by a simple estimate of the optical depth due to dust along the line of sight that \lyman\ photons suffer from. Averaging over the (face-on) lines of sight from young stellar cluster which we consider a source of \lyman\ radiation, we find a mean optical depth of $\tau_D=17$, albeit with a wide distribution skewed to small values (for example, 40\% of the emitters have $\tau_D<5$). The optical depth from dust in the source cells alone is about 8 on average, again with a large scatter.

These two experiments show that \textit{to transform \althaea\ into an object with significant transmission of \lyman a selective depletion of the dust contained in \HII\ regions is required.} Quite naturally, one may wonder whether the dust-free bubbles model is based on some physical grounds. While it is plausible that \HII\ regions near young stellar clusters might be dust-depleted to some extent \citep[e.g. through dust destruction in the ionized regions, e.g.][]{Pavlyuchenkov2013}, we cannot uniquely individuate a physical mechanism that completely destroys the dust in the vicinity of the star-forming regions. 
However, it is worth mentioning the results by \cite{Draine2011} who first studied radiation pressure effects on the internal density structure of \HII\ regions. He finds that if the \HII\ region is powered by stars with a strong ionizing flux surrounded by an initially dense medium, the gas and dust in the cavity get evacuated and pile up into a shell marking the ionized boundary. 

This effect is not fully caught by our simulations. In spite of their high resolution ($<30$ pc), our simulations might be insufficient to properly describe both the \lyman\ \RT\ in the inherently turbulent medium characterizing molecular cloud interiors, and the above radiative feedback effects: In order to do so, sub-pc resolution would be required.

\begin{figure}
\centering
\includegraphics[width=0.49\textwidth]{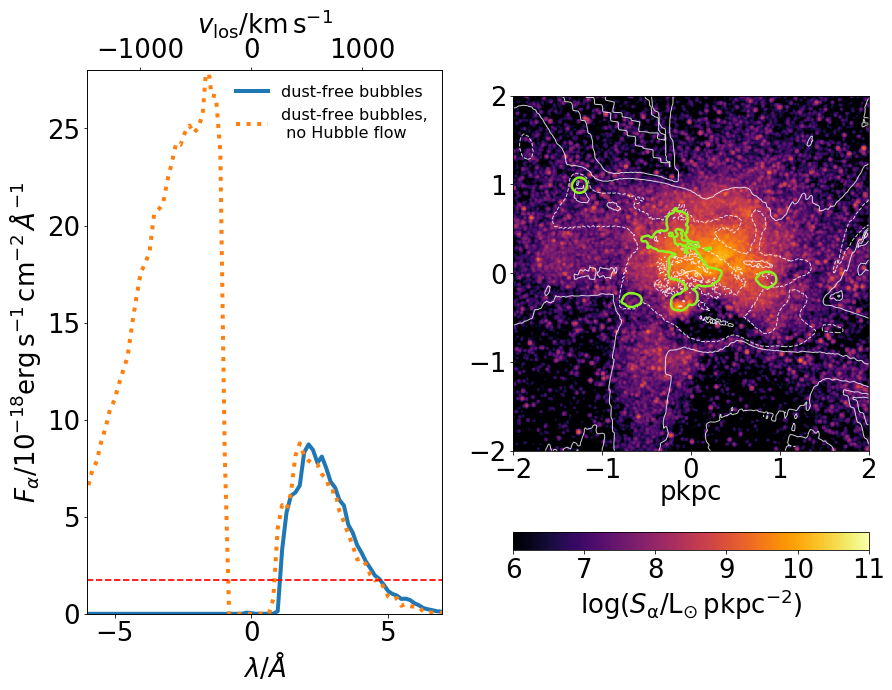}
\caption{\lyman\ spectrum (left) and morphology (right) for the dust-free bubbles model along the los with a large EW (22 \AA{}). For comparison, we also show the spectrum obtained from artificially switching off the Hubble flow (dotted). The dashed line indicates the UV continuum level. The contours in the right panel show the gas column density (thin, white lines) and the intrinsic \lyman\ emission (thick, green lines).
\label{fig_lae}
}
\end{figure}

With the above caveats, we compare in more detail the fiducial and the dust-free bubbles model. For the dust-free bubbles, we have run simulations with 48 lines of sights. In these simulations, we find qualitatively similar dependencies of the \lyman\ escape on inclination as for the fiducial model. However, the \lyman\ and UV luminosities are higher in the dust-free bubbles model (see Fig. \ref{fig_lae}), with \EW\ reaching 22 \AA{}. In this case, \althaea\ would be classified as a \LAE\ by the \EW\ cuts typically adopted by observational practice ($\sim 20$ \AA{}). The maximum escape fraction is $f_\alpha=0.15$. For one los with a high \EW, we show in Fig. \ref{fig_lae}  both the resulting spectrum (left panel) and the morphology (right panel) of the escaping \lyman\ radiation; we indicate the continuum flux level with a dotted line. The line widths in the dust-free case are similar to the fiducial case; for the lines of sight with EW > 20 \AA{}, we find a width of $\approx 600$ km/s. In the dust-free bubbles model, the total Ly$\alpha$ flux predicted by our model ($\sim 2\times 10^{-17}$ erg s$^{-1}$ cm$^-2$, for a FWHM$\sim$3 \AA{}) is consistent with the ones typically observed in $z\sim 7$ LAEs \cite[e.g.][]{Vanzella2011,Caruana2014,Debarros2017}, thus being detectable in $\sim$ 15 hr with the FORS2 spectrograph on the ESO Very Large Telescope.

\begin{figure}
\centering
\includegraphics[width=0.49\textwidth]{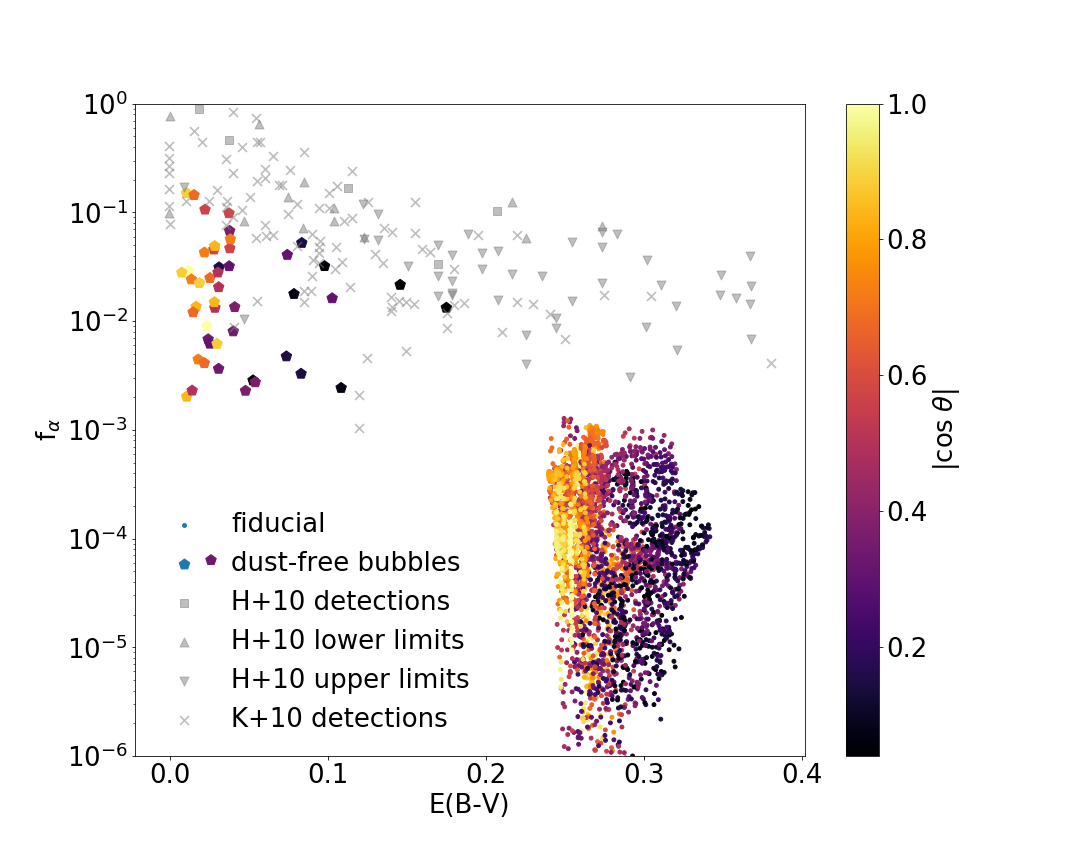}
\caption{\lyman\ escape fraction $f_\alpha$ as a function of extinction for each los of two simulation sets, compared with the data from \protect\cite{Hayes2010,Kornei2010} as compiled by \protect\cite{Hayes2011}. The shading indicates the inclination of the respective los, with a value of 0 (1) indicating edge-on (face-on) orientation.
\label{fig_ebminusv}
}
\end{figure}

In Fig. \ref{fig_ebminusv} we report the relation between $f_\alpha$ and $E\rm (B-V)$ for both the fiducial and dust-free bubbles models at $z=7.2$\footnote{To calculate $E\rm (B-V)$, we used the optical depth in the V/B-band at 0.552/0.442 $\mu \mathrm{m}$ as calculated in the UV continuum RT.}. The main effect of the  dust-free bubbles model is to shift horizontally rightwards the points in the plot. 

For comparison, in Fig. \ref{fig_ebminusv} we also plot available observational data from \cite{Kornei2010,Hayes2011} for low-redshift galaxies ($z=2-3$). The lower (upper) limits taken from \cite{Hayes2011} correspond to objects detected in \lyman\ (H$_\alpha$), but not in H$_\alpha$ (\lyman).

\begin{figure}
\centering
\includegraphics[width=0.49\textwidth]{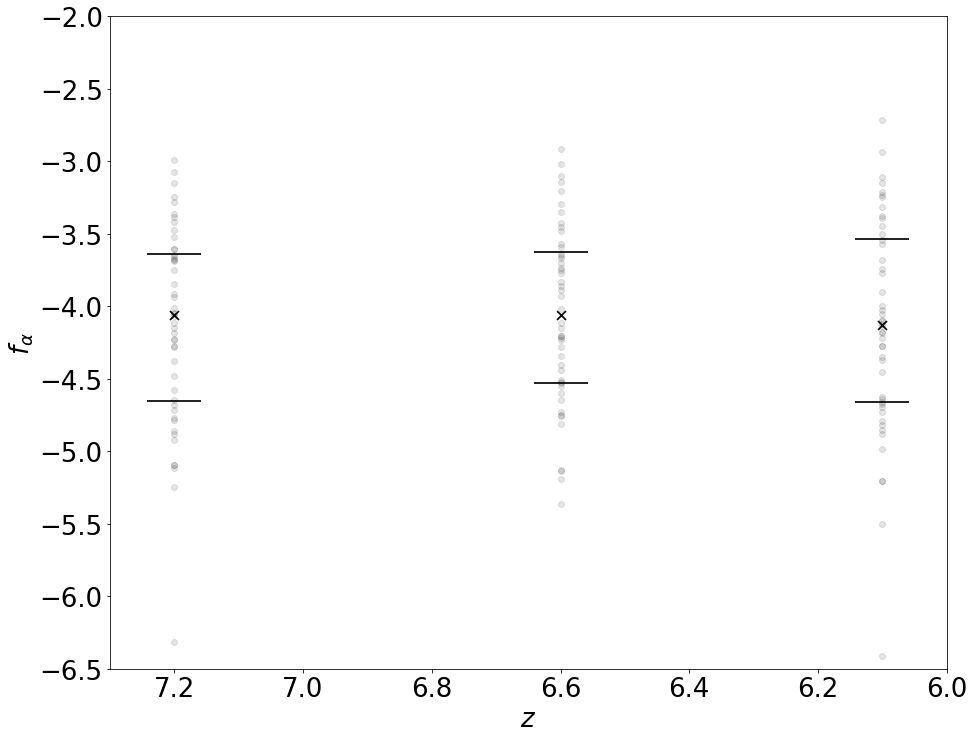}
\caption{The escape fraction of \lyman\ radiation, $f_\alpha$, as a function of redshift. We also show the median (crosses) of the 48 lines of sight simulated for each redshift. Bars indicate the lower/upper quartiles. 
\label{fig_redshifts}
}
\end{figure}

While in this paper, we focused on \althaea\ at redshift $z=7.2$, we show the escape fraction $f_\alpha$ for three different evolutionary stages and 48 lines of sight per stage in Fig. \ref{fig_redshifts} together with the median (crosses) and the lower/upper quartiles (bars). At the additional redshifts of 6.5 and 6.1, \althaea\ has a stellar mass of 1.4 and 2.0 $\times$ $10^{10} \msun$, respectively. We find only a mild variations from snapshot to snapshot in terms of the escape fraction; interestingly, the extinction of the UV is the highest at $z=7.2$, whereas the total dust mass is the lowest at this evolutionary stage. This can be related to the fact that in this stage, recent burst of star formation have deposited metals in their immediate surroundings, leading to large attenuation.

\subsubsection*{Summary}

We conclude that the clumpy distribution of dust is the key factor inducing \lyman\ line quenching, with the peculiar velocity field playing a sub-dominant role.  In turn, the dust is clumpy because grains tend to cluster around their production factories, at these high redshifts predominantly massive stars \citep{Todini2001}.  

\begin{figure}
\centering
\includegraphics[width=0.49\textwidth]{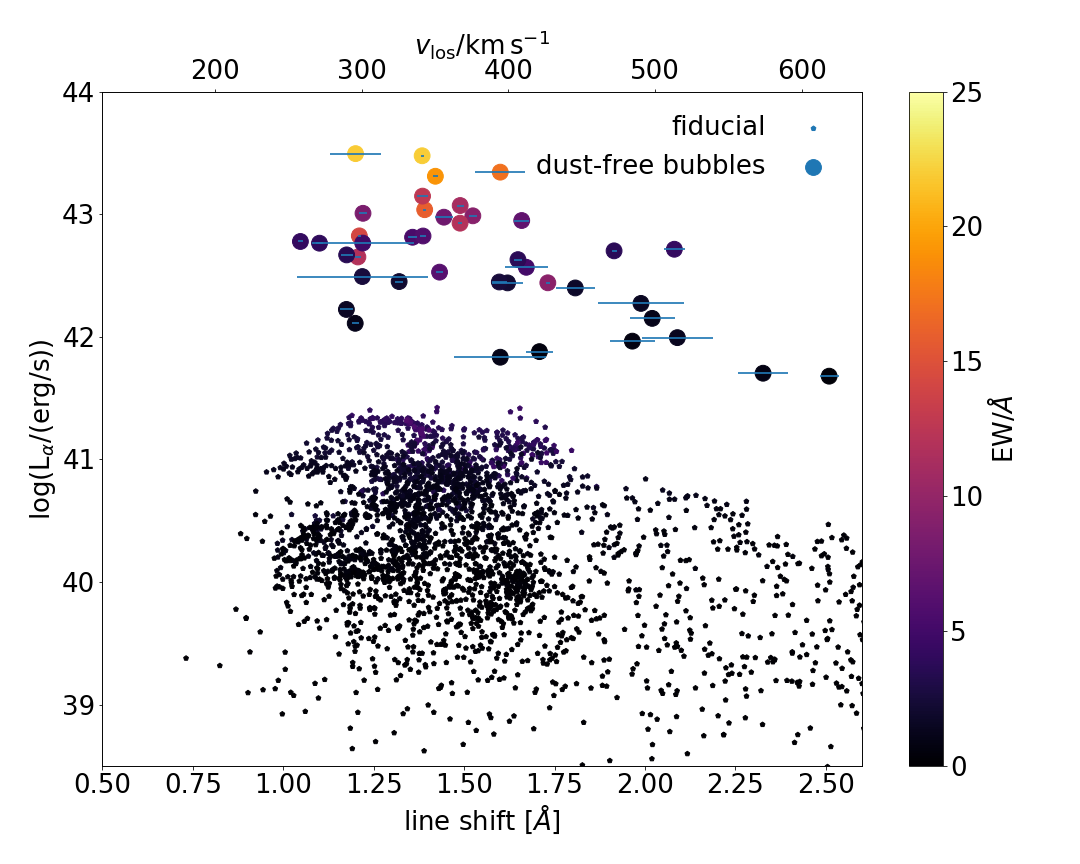}
\caption{\lyman--\CII\ line shift vs. \lyman\ luminosity for the fiducial (circles) and the dust-free bubbles model (pentagons) for different lines of sight. Errorbars are the errors of the fits to the \lyman\ line profiles. The x-axis on top shows the line shifts in units of velocity along the line of sight. For clarity, we have omitted the error bars for the fiducial sample. The average error is $0.1 \pm 0.1$ \AA{}.
\label{fig_cii_shift} }
\end{figure}  

\subsection{\lyman--[CII] line shift}

In Fig. \ref{fig_cii_shift}, we show the relative shift between the \lyman\ and [CII] lines as a function of the \lyman\ luminosity for different los and the fiducial/dust-free bubbles model. Within each of the two samples, we find a weak anti-correlation, i.e. the shift increases towards lower \lyman\ luminosities (and \EW). In our case, this is readily explained by the low escape fractions. As already discussed in the literature \citep[see e.g.][]{Laursen2009}, \lyman\ photons far in the wings are more subject to absorption, since they typically have traveled a longer path through the dusty medium; they originate from regions of high gas densities, and undergo a large number of scatterings, which in turn induce changes of direction and enhance their path lengths. Enhanced pathlengths boost the attenuation of these photons, thus dust damps preferentially the outer parts of the spectrum, even though the dust absorption cross section itself is only weakly depending on frequency. Accordingly, we find the average number of scatterings prior to escape to be anti-correlated with the observed luminosity along a given los.

\cite{Verhamme2018} reported a positive correlation between line shift and \lyman\ line width. We do not find such a relation; this holds also in the dust-free bubbles case, i.e. when \althaea~is a \LAE. However, we note that even in this favorable case, the escape fraction is at the percent level due to severe dust attenuation. This suggests the relationship might only valid in low-attenuation systems like the ones \cite{Verhamme2018} consider, where the escape is largely driven by the outflow of gas. In contrast, in our case, it is the dust content that dominates the escape, and outflows only play a minor role.

\section{Summary and Conclusions}\label{sec_discussion}

Performing radiative transfer, we have post-processed \althaea, a simulated prototypical \LBG\ at $z=7.2$ \citep{Pallottini2017}, to obtain its \lyman , continuum, and \CII\ line properties.
We used the publicly available dust continuum code \code{Skirt} \citep{Baes2003,Camps2015}, a semi-analytic model for the \CII\ line \citep{vallini:2015,pallottini2017a}, and \code{Iltis}, an updated version of the \lyman\ code used by \cite{Behrens2018a,Behrens2014a,Behrens2014b}.
We included both the interaction of radiation with gas and with dust, and employed a simple model for \HII\ bubbles forming around star-forming regions, as we do not directly simulate ionizing radiation.
We performed our simulations using up to 3072 lines of sight. 

In our fiducial simulation, \lyman\ radiation escapes \althaea\ solely from one side and one quadrant on the outskirt of the disk, characterized by a low column density gap in between two filamentary structures. Both the star-forming core and the $\sim$ 100 pc-scale star-forming clumps remain dark in \lyman.
\althaea\ is a resilient \LBG\ with low \lyman\ \EW s (<6 \AA{}) compared to an intrinsic Lya \EW\ of 103 \AA{}; the implied escape fractions  are less than 0.1\%.  

In contrast to studies investigating highly idealized, isolated disk galaxies \citep{Verhamme2012,Behrens2014b}, we only find a weak correlation between inclination of the los and the escape fraction/\lyman\ luminosity because of the complex, anisotropic structure of gas/dust in and around \althaea. However, the escape fraction $f_\alpha$ strongly depends on the specific los, with variations of up to six orders of magnitude.

The emerging spectra typically show a well-defined asymmetric, single-peaked shape located redwards of the \lyman\ line center. However, the spectrum prior to entering the \IGM\ is double-peaked. The dominant blue peak is washed out by the IGM, as bluer photons are progressively redshifted into resonance and scatter on residual hydrogen out of the line of sight. 
We have investigated the cause for the very small escape fraction found in our fiducial simulation, and find that transfer is dominated by the clumpiness of dust close to star-forming regions, absorbing \lyman\ photons very efficiently. Outflows do not generally play a major role for the escape mechanism, apart for some specific los.

To turn our \LBG\ into a \LAE, we have artificially removed dust from the galaxy. While reducing the overall dust content does not make \althaea\ a LAE, by selectively removing the dust content from star-forming regions we reach \EW s up to 22 \AA{} and escape fractions of up to $f_\alpha$=15\%. This further justifies our conclusion that it is the dust within the star-forming clumps that dominates the \RT. As a by-product, neither the precise value for the dust-to-metal ratio, nor the inclusion/exclusion of ionized regions around stellar clusters change our results dramatically.
While in this paper we have mostly focused on one evolutionary stage of \althaea, we have also run similar simulations on two snapshots at different redshifts that exhibit lower star formation rates, without dramatic changes in the escape fraction. 

Taken at face-value, our results raise the question of what mechanism could turn \LBG s into \LAEs\ and vice versa in the popular duty-cycle scenario. Such a mechanism would require to lower the attenuation by orders of magnitude within a relatively short time. Stellar clusters that are less attenuated in our simulations have typically moved out of their birth clouds, and are therefore too old to contribute to the ionizing flux generating \lyman. A stronger feedback mechanism dispersing the molecular clouds more quickly, however, might be a viable solution if it can also get rid of the dust.

Finally, we investigated the relation between the \lyman\ luminosity and the line shift of the \lyman\ line, measured with respect to the [CII] line. For the face-on fiducial model, the \lyman\ shift with respect to the [CII] line is 1 \AA{} ($\sim 250$ km/s). We find a negative correlation between the two, i.e. the shift increases towards lower \lyman\ luminosities (and \EW). This does not support the suggestion by \cite{Pentericci2016} that smaller line shifts might correlate with a more strongly damped \lyman\ line due to an enhanced attenuation in the \IGM. This difference might be related to the overall small value of $f_\alpha$ in our case.

Since our simple model of ionized bubbles around stellar clusters can only be considered a rough estimate, direct \RT\ of the ionizing photons emerging from stars will be required to get the full picture. Another caveat of our work is the lack of substructure of molecular clouds in our simulation due to our resolution limit of 25 pc. Sub-grid models for the escape of photons from the multiphase molecular clouds will be necessary \citep[also see the work by][]{Hansen2006,Gronke2016,Kimm2019}, including e.g. the evaporation of the cloud by young stars. We plan to address these issues in future work. 

\section*{Acknowledgements}
AF acknowledges support from the ERC Advanced Grant INTERSTELLAR H2020/740120. 
LV acknowledges funding from the European Union's Horizon 2020 research and innovation program under the Marie
Sk\l{}odowska-Curie grant agreement No. 746119.
We acknowledge use of the Python programming language \citep{VanRossum1991}, and use of the packages IPython \citep{Perez2007}, Matplotlib \citep{Hunter2007}, Numpy \citep{VanDerWalt2011}, Pymses \citep{Labadens2012}, and Jupyter \citep{LoizidesSchmidt2016}.
This research made use of Astropy, a community-developed core Python package for Astronomy \citep{astropy}.




\bibliographystyle{mnras}
\bibliography{references,packages} 



\appendix

\section{Details on the \lyman RT}\label{sec_app_rt}

\subsection{Peeling-off algorithm}

As presented in Sec. \ref{sec_rt}, the peeling-off algorithm requires to integrate the optical depth along the los to each scattering event. In order to speed up this calculation, we discard contributions whenever the line of sight optical depth exceeds $\tau_o = 20$; since the weight of each contribution scales exponentially, this means we discard contributions that add less than $O(10^{-9})$, which translates to $10^{33}$ erg/s in our units.

\subsection{Hubble flow}

Photons redshift in-between scatterings. In order to make sure we resolve the passage of a photon on the blue side of the spectrum through the line center, we invoke a limit on the maximum step size of $\sim 1$ kpc, which corresponds to a velocity shift of 0.9 km/s at $z=7.2$.

\subsection{Dust properties}

For the \lyman, we used the same dust model as in \code{skirt}, that is, the \cite{Weingartner2001} model. However, as we are interested in the line only, we assumed the cross sections, the albedo, and the asymmetry parameter to be the constant with respect to frequency. Their values changes only modestly (e.g., the cross section varies by $\sim 3\%$ within $\pm 10\;\AA{}$ of the \lyman\ line center), and thus we chose to set them to their values at the \lyman\ line center.

We explicitly checked that UV continuum RT in \code{skirt} and \lyman\ RT in \code{Iltis} are consistent with each other, that is, we checked that the \lyman\ RT and the UV continuum RT yield the same results at the \lyman\ line center if we only consider dust (by removing all gas artifically).

\subsection{Parameters of the RT}

We launch \lyman\ photons from each stellar cluster that has an intrinsic \lyman luminosity of above 10$^{38}$ erg/s in order to reduce the number of sources we need to consider. The intrinsic spectrum is a Gaussian with a width of 10 km/s; As we expect our \lyman\ photons to originate from recombinations in the vicinity of young stars, the width is motivated from the typical velocity dispersion in such regions.  We arrive at a total number of 2590 sources. We launch a minimum of $10^4$ ($5 \times 10^3$ for the runs using 48 lines of sight) tracer photons (equivalently; photon packages) per source. While this number is small given the fact that the escape fraction is very low, we acquire of the order $O(10^6)$ contributions from the peeling-off scheme and find good convergence. For the runs with 3072 lines of sight, we only launch $10^3$ photon packages in the \lya\ transfer. We have verified that these runs have sufficiently converged using the simulations with a lower number of lines of sight.  We make use of the usual acceleration scheme, skipping core scatterings by cutting the thermal distributions of scattering atoms if the dimensionless frequency of the photon is $x<3$\cite[e.g.][]{Dijkstra2006}. For the continuum photons, we use $10^6$ photon packages, as this value is sufficient to achieve convergence.

\bsp	
\label{lastpage}
\end{document}

%% file: package.tex
\usepackage{mathptmx}
\usepackage{txfonts}

\usepackage[T1]{fontenc}
\usepackage{ae,aecompl}

\usepackage{graphicx}	
\usepackage{amssymb}	
\usepackage{mathrsfs}   

\usepackage{ifthen}

%% file: definitions.tex
\newcommand\code[1]																																																																																								{\textsc{\MakeLowercase{#1}}}
\newcommand{\quotes}[1]{``#1''}
\def\althaea{Alth{\ae}a}
\def\lyman{Ly$\alpha$}
\def\CII{\hbox{[C~$\scriptstyle\rm II $]}}
\def\HII{\hbox{H$\scriptstyle\rm II $}}
\def\lya{Ly$\alpha$\,}
\def\CIIion{\hbox{C~$\scriptstyle\rm II $}}

\def\msun{{\rm M}_{\odot}}
\def\lsun{{\rm L}_{\odot}}

\def\myr{{\rm Myr}}

\def\msunyr{\msun/{\rm yr}}

\def\cold{{\rm cm}^{-2}}
\def\kms{{\rm km}/{\rm s}}
\def\surfd{\msun/{\rm kpc}^{-2}}
\def\surfl{\lsun/{\rm kpc}^{-2}}

%% file: abbreviation.tex
\newcounter{SEDDone}
\def\SED{\ifthenelse{\equal{\arabic{SEDDone}}{0}}{Spectral Energy Distribution (SED)\setcounter{SEDDone}{1}}{SED}}
\newcounter{EWDone}
\def\EW{\ifthenelse{\equal{\arabic{EWDone}}{0}}{equivalent width (EW)\setcounter{EWDone}{1}}{EW}}
\newcounter{IMFDone}
\def\IMF{\ifthenelse{\equal{\arabic{IMFDone}}{0}}{Initial Mass Function (IMF)\setcounter{IMFDone}{1}}{IMF}}
\newcounter{ISMDone}
\def\ISM{\ifthenelse{\equal{\arabic{ISMDone}}{0}}{interstellar medium (ISM)\setcounter{ISMDone}{1}}{ISM}}
\newcounter{IGMDone}
\def\IGM{\ifthenelse{\equal{\arabic{IGMDone}}{0}}{intergalactic medium (IGM)\setcounter{IGMDone}{1}}{IGM}}
\newcounter{RTDone}
\def\RT{\ifthenelse{\equal{\arabic{RTDone}}{0}}{radiative transfer (RT)\setcounter{RTDone}{1}}{RT}}
\newcounter{LAEDone}
\def\LAE{\ifthenelse{\equal{\arabic{LAEDone}}{0}}{Lyman-$\alpha$ emitter (LAE)\setcounter{LAEDone}{1}}{LAE}}
\def\LAEs{\ifthenelse{\equal{\arabic{LAEDone}}{0}}{Lyman-$\alpha$ emitters (LAEs)\setcounter{LAEDone}{1}}{LAEs}}
\newcounter{LBGDone}
\def\LBG{\ifthenelse{\equal{\arabic{LBGDone}}{0}}{Lyman break galaxy (LBG)\setcounter{LBGDone}{1}}{LBG}}
\def\LBGs{\ifthenelse{\equal{\arabic{LBGDone}}{0}}{Lyman break galaxies (LBGs)\setcounter{LBGDone}{1}}{LBG}}
\newcounter{FIRDone}
\def\FIR{\ifthenelse{\equal{\arabic{FIRDone}}{0}}{Far Infra-Red (FIR)\setcounter{FIRDone}{1}}{FIR}}